\documentclass[12pt]{article}

\RequirePackage{caption2}
\RequirePackage{graphicx}

\begin{document}

\begin{center}
{\Large {\bf Results of a search for 2$\beta$-decay of $^{136}$Xe
with high-pressure copper proportional counters in Baksan Neutrino
Observatory}}

\vskip 0.4cm

{\bf Ju.M.~Gavriljuk$^{a}$, A.M.~Gangapshev$^{a,}$\footnote{Talk at
5th International Conference on Non-Accelerator New Physics
(NANP'05), Dubna, Russia, 20--25 June 2005.}, V.V.~Kuzminov$^a$,
\\S.I.~Panasenko$^b$, S.S.~Ratkevich$^{b}$}

{$^a$~\it Baksan Neutrino Observatory, INR RAS, Russia}

{$^b$~\it Karazin Kharkiv National University, Ukraine}

\end{center}

\begin{abstract}
The experiment for the 2$\beta$-decay of $^{136}$Xe search with two
high-pressure copper proportional counters has been held in Baksan
neutrino observatory. The search for the process is based on
comparison of spectra measured with natural and enriched xenon. No
evidence has been found for 2$\beta$(2$\nu$)- and
2$\beta$(0$\nu$)-decay. The decay half lifetime limit based on data
measured during 8000 h is T$_{1/2}$$\geq$8.5$\cdot$10$^{21}$yr for
2$\nu$-mode and T$_{1/2}$$\geq$3.1$\cdot$10$^{23}$yr for 0$\nu$-mode
(90$\%$C.L.).
\end{abstract}

\begin{center}
\bf Introduction
\end{center}

The experimental investigation of the 2$\beta$-decay of $^{136}$Xe
has been started more then 20 years ago. But till now both two
neutrino and neutrinoless modes of this process are not observed
yet. The results of last experiments are presented in
tab.\ref{tab1}. The theoretical estimations of half lifetime for
$2\beta(2\nu)$-decay are presented in tab.\ref{tab2}. It is
necessary to mention that in \cite{r4} only one spectrum (measured
with enriched $^{136}$Xe) was obtained. To calculate their limit it
was assumed that at any effect/background ratio in the energy range
under investigation the effect did not exceed the actually measured
background increased by a systematic error given in $\sigma$ units
($\sigma$ is a standard  deviation). Actually, in the case of search
of 2$\beta$(2$\nu$)-decay, this method does not allow one to find
the effect and could be used only to find a limit. To discover such
an effect it is necessary either to measure directly the background
of the installation under the same conditions or to simulate the
background. In DAMA/LXe experiment such work has not been done. In
our work the measurements were performed with enriched xenon (93$\%$
of $^{136}$Xe) and natural xenon simultaneously.

\begin{center}
\bf Experimental setup
\end{center}

\begin{figure*}[hp]
\begin{center}
\setcaptionmargin{5mm} \onelinecaptionstrue
\includegraphics[width=10.5cm,angle=0.]{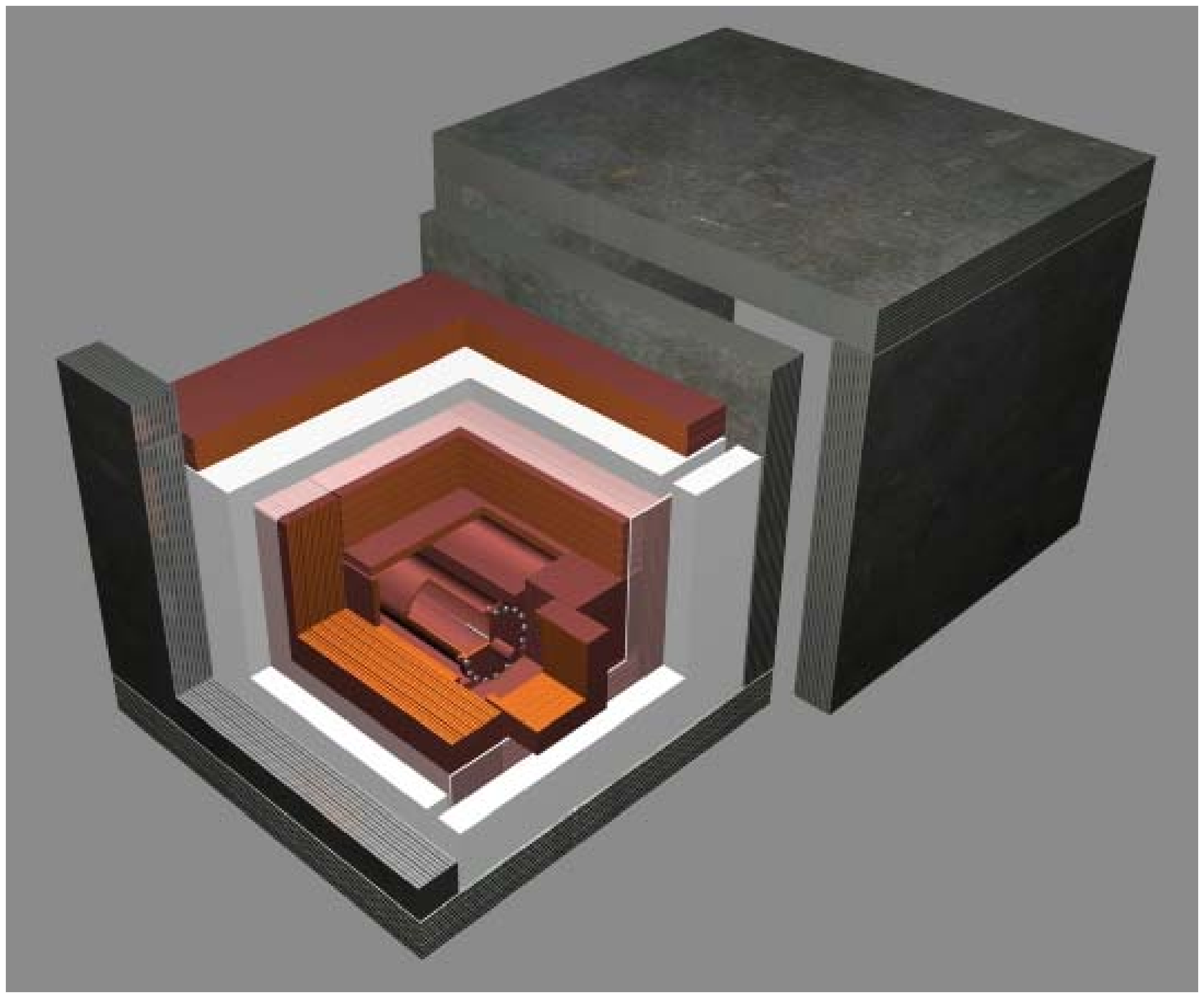}
\captionstyle{normal} \captionstyle{flushleft}\caption{\label{inst}
The view of the installation.}
\end{center}
\end{figure*}

\begin{figure*}[t!]
\begin{center}
\setcaptionmargin{5mm} \onelinecaptionstrue
\includegraphics[width=15.5cm,angle=0.]{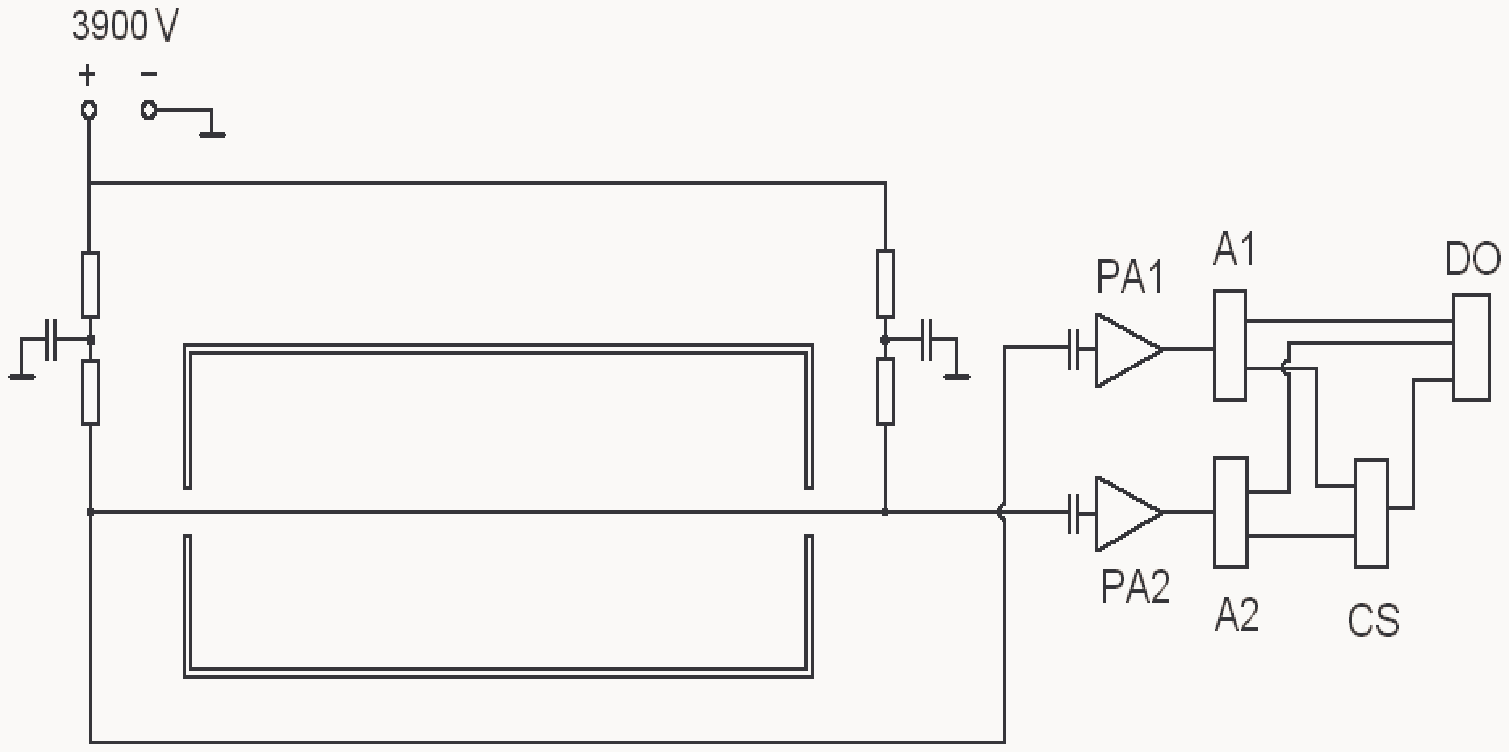}
\captionstyle{normal}
\captionstyle{flushleft}\caption{\label{el_sch} The electric scheme
of installation. PA1 and PA2 - preamplifiers, A1 and A2 -
amplifiers, CS - coincidence scheme, DO - digital oscilloscope.}
\end{center}
\end{figure*}

\begin{figure*}[t!]
\begin{center}
\setcaptionmargin{5mm} \onelinecaptionstrue
\includegraphics[width=7.5cm,angle=270.]{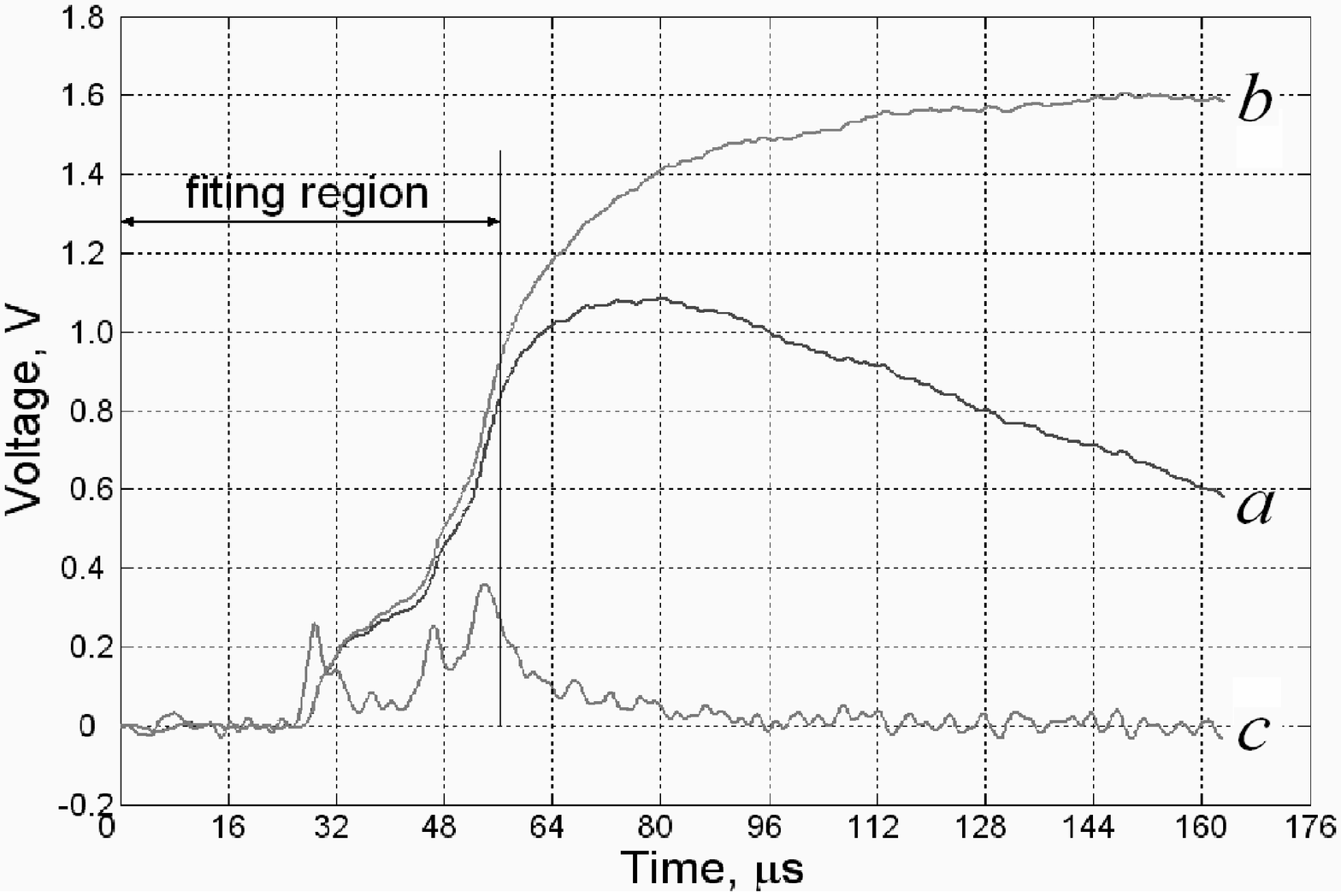}
\captionstyle{normal} \caption{\label{pulse} The samples of the
pulses ($a$-initial pulse, $b$-reconstructed pulse, $c$-differential
of reconstructed pulse).}
\end{center}
\end{figure*}

The measurement was carried out with two copper proportional
counters (CPC $N1$ and CPC $N2$). When one of them was filled with
enriched xenon, the other one was filled with natural xenon with
extracted light isotopes (9.2$\%$ of $^{136}$Xe). Both CPC-s are
surrounded by passive shield consisting of 20 cm of copper, 8 cm of
borated polyethylene and 15 cm of lead (see fig.\ref{inst}). The
installation is located in the deep underground laboratory of the
BNO INR RAS at the depth 4900 m w.e, where the flux of muons is
decreased by factor 10$^7$ and evaluated as $2.23\cdot10^{-9}$
cm$^{-2}s^{-1}$ \cite{r9}. The parameters of the CPC-s are: working
pressure - 14.8 $at$, fiducial volume - 9.16 l, applied voltage -
3800 V. Signals PC1 and PC2 were read out from both ends of the
anode wire through preamplifiers. Then they were supplied to the
digital oscilloscope through amplifiers. (see fig.\ref{el_sch}).
Such a scheme of read out system allowed us to determine a relative
coordinate ($\beta$) of the events along the anode wire by the
following equation $\beta=A1/(A1+A2)$, where A1 and A2 amplitudes of
pulses PC1 and PC2. Parameter $\beta$ is used to reject both events
of microdischarges in the outward high voltage circuits and those on
the anode insulator surfaces. To exclude the influence of capacity
charge decay on the pulse amplitude each pulse is reconstructed by
software taking into account this decay. The value of reconstructed
pulse at the point of initial pulse maximum was used as a amplitude
to construct the energy spectrum. For detailed analysis the
following pulse shape parameters was used: the pulse rise time
($\tau$) and parameter $\delta$ defined as
\begin{eqnarray*}
\delta=100\cdot(1-\frac{\sum{(y_{i}-g_{i})^{2}}}
{\sum{(y_{i}-\overline{y})^{2}}}).
\end{eqnarray*}
Where the $y_{i}$ is value of differential of reconstructed pulse at
the point {\it i}, $g_{i}$ - value of gaussian at the same point,
$\overline{y}$ - mean value of the differential of pulse in fitting
region. The region of fitting by gaussian of the differential of
pulse is restricted by point of pulse beginning and point where the
initial pulse has a value $0.9\cdot A$. In fig.\ref{pulse} samples
of the pulses are shown.

\begin{center} \bf Results \end{center}

From previous experiment \cite{r3} it was seen that significant part
of the counters background is due to $\alpha$-particles produced in
gas volume. To define parameters of $\alpha$-particle events the
measurement of background of one CPC-s filled with xenon with
admixtures of $^{222}$Rn ($\alpha$-particles with energy of 5.49
MeV, 6.02 Mev and 7.69 Mev) was done. The results of this
measurement are presented in \cite{r10}. To define parameters of
events from $\gamma$-rays the measurement with $^{232}$Th-source was
carried out. Before measurements with $^{222}$Rn and $^{232}$Th and
during main measurements the calibration of both CPC-s by
$^{22}$Na-source was done. The energy resolution was determined as
13.5$\%$, 9.5$\%$ and 8.1$\%$ for 511 keV, 1275 keV and 1592 keV,
respectively.

Main measurements consist of 5 runs. In the first run CPC $N1$ was
filled with enriched xenon and CPC $N2$ with natural xenon. Duration
of the run was $\sim$3000 h. Then first CPC was refilled with
natural xenon, second one with enriched xenon, after each run the
refilling procedure was repeated. Such a procedure allows
elimination of systematic error from possible differences between
counters. For analysis the even number of runs was used. To exclude
the contribution of $^{222}$Rn (which goes from gas cleaning system)
to the CPC-s background the data of the first 500 h of measurements
were not used for analysis. During this time radon decays almost
completely. The calibration of the counters was carried out every 2
weeks (~300 h).

\begin{figure*}[t!]
\begin{center}
\setcaptionmargin{5mm} \onelinecaptionstrue
\includegraphics[width=11.5cm,angle=0.]{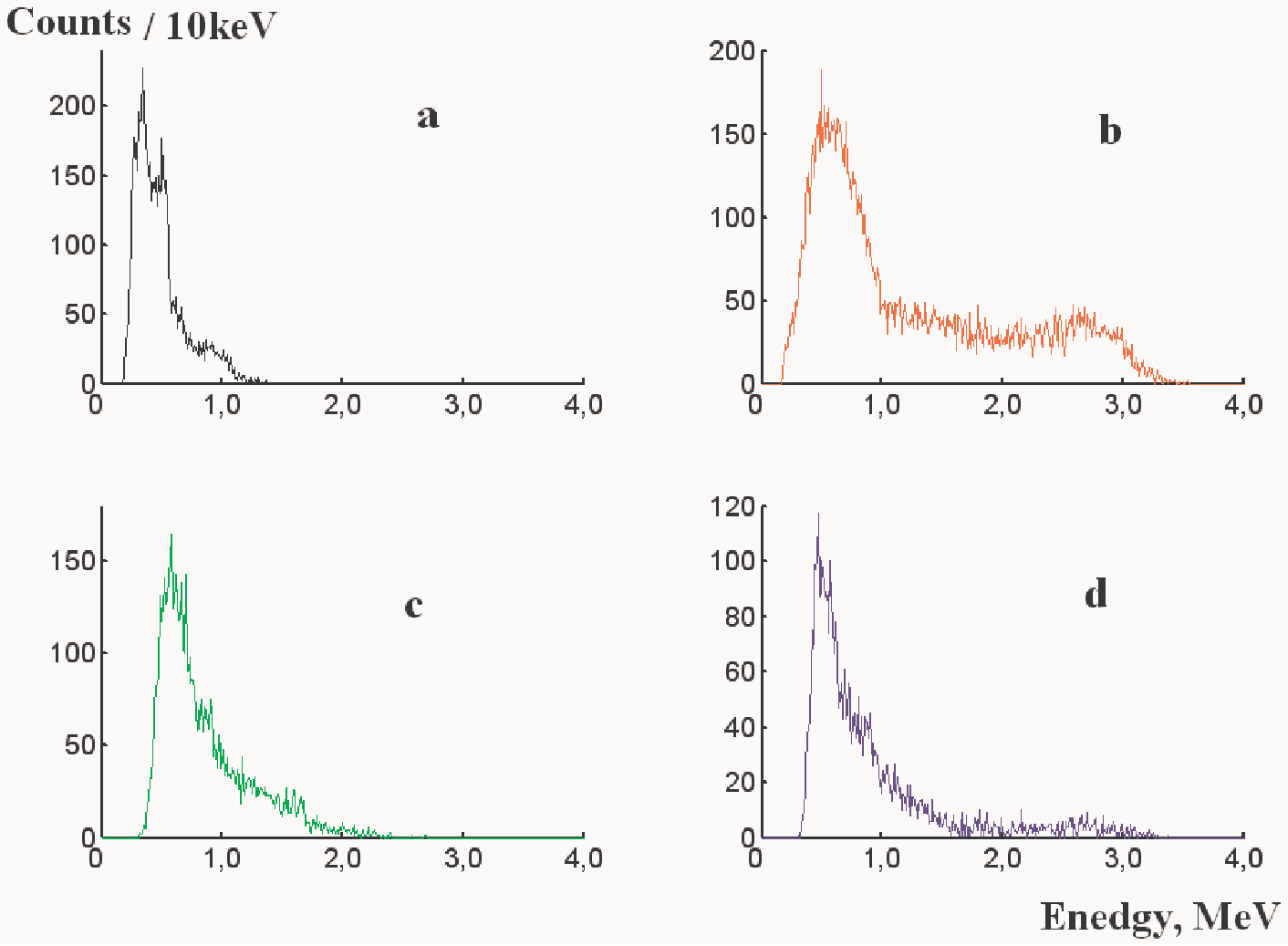}
\captionstyle{normal} \captionstyle{flushleft}\caption{\label{n_e}
The energy spectra: {\bf a} - $^{22}$Na, {\bf b} - $^{222}$Rn, {\bf
c} - $^{232}$Th and {\bf d} - background (2000 h).}
\end{center}
\end{figure*}

\begin{figure*}[t!]
\begin{center}
\setcaptionmargin{5mm}
\onelinecaptionstrue
\includegraphics[width=13.5cm,angle=0.]{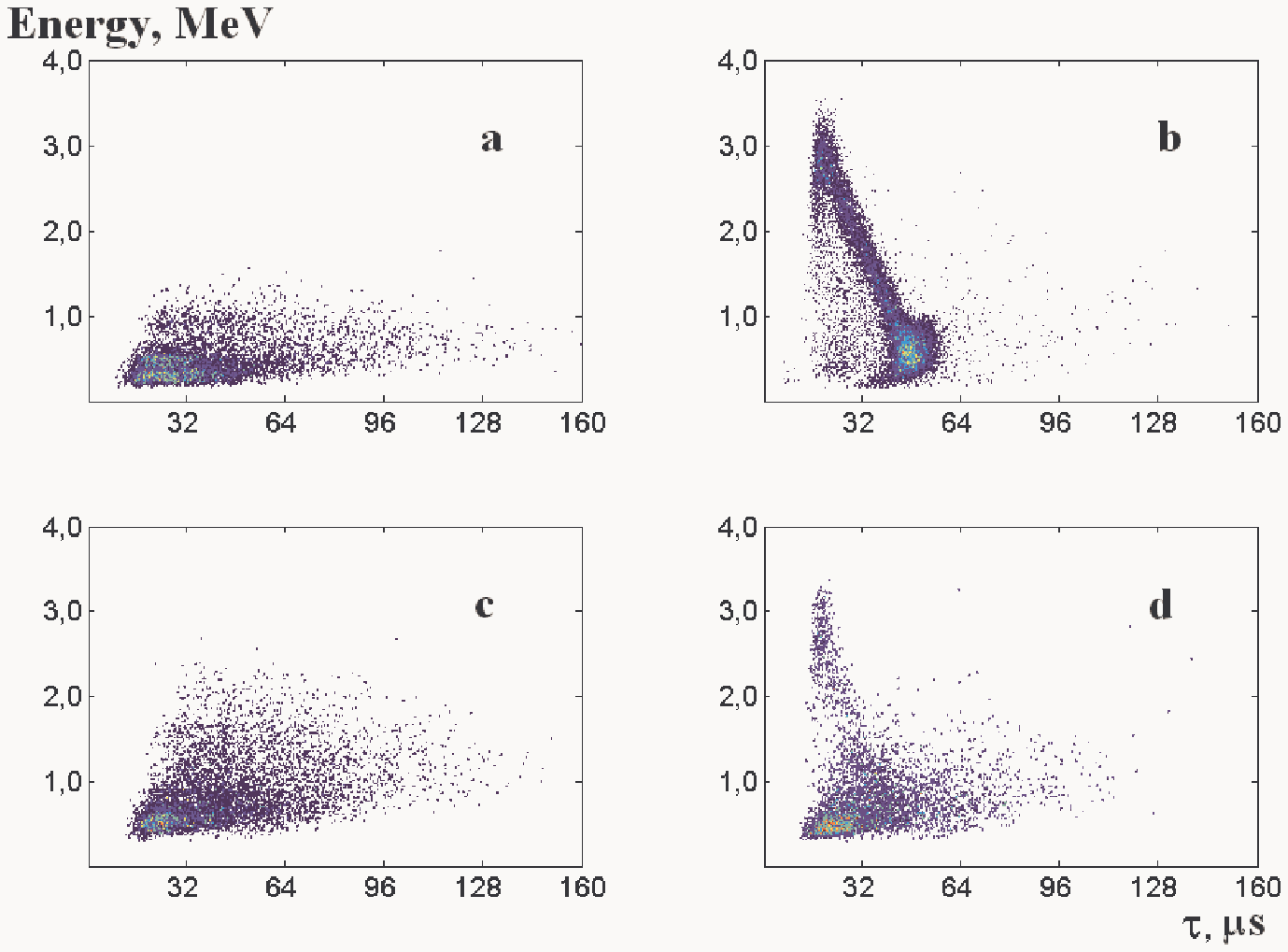}
\captionstyle{normal} \captionstyle{flushleft}\caption{\label{e_t}
Distribution of events versus energy and $\tau$, {\bf a} -
$^{22}$Na, {\bf b} - $^{222}$Rn, {\bf c} - $^{232}$Th and {\bf d} -
background (2000 h).}
\end{center}
\end{figure*}

\begin{figure*}[t!]
\begin{center}
\setcaptionmargin{5mm}
\onelinecaptionstrue
\includegraphics[width=13.5cm,angle=0.]{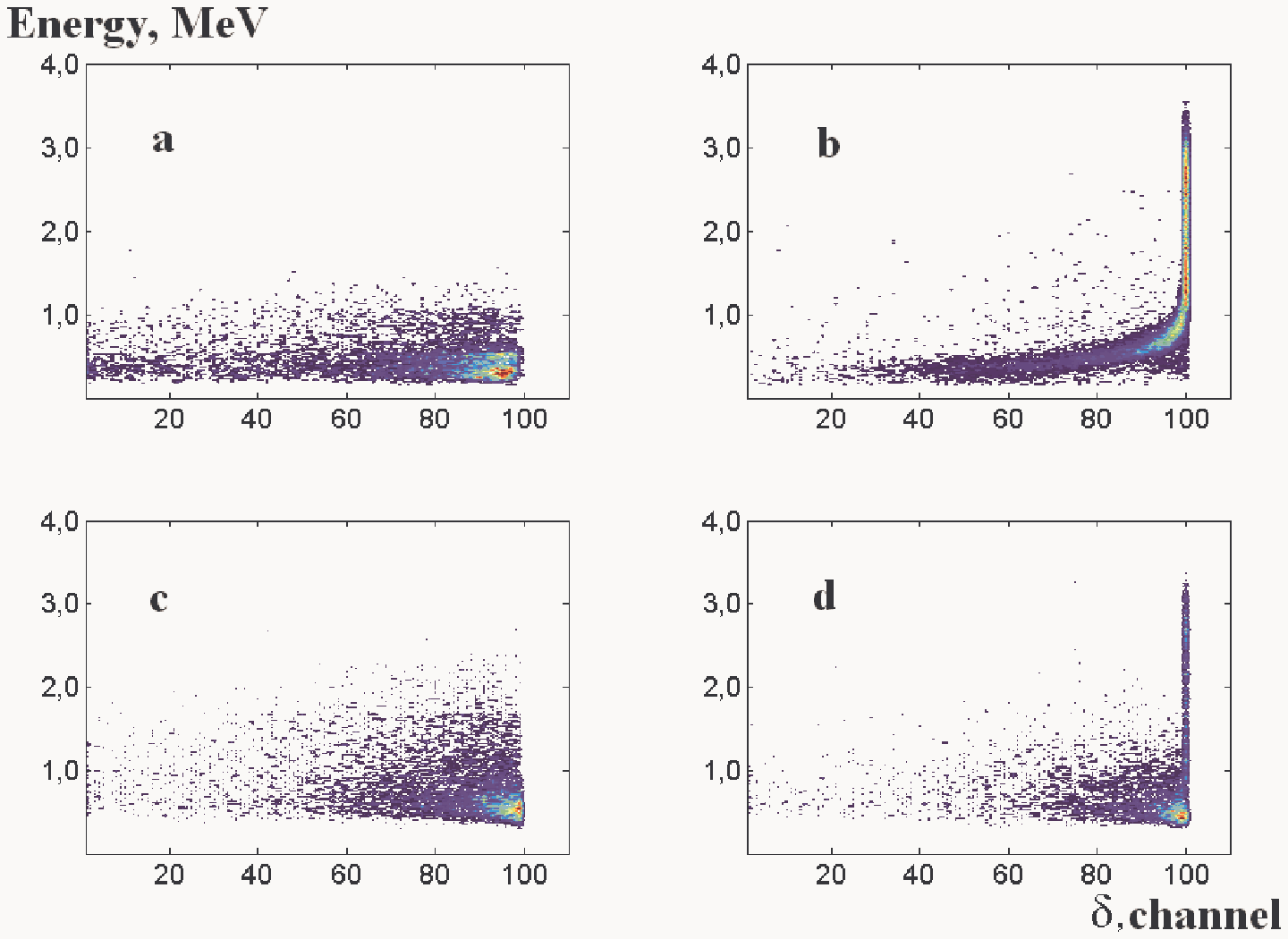}
\captionstyle{normal} \captionstyle{flushleft}\caption{\label{e_p}
Distribution of events versus energy and $\delta$: {\bf a} -
$^{22}$Na, {\bf b} - $^{222}$Rn, {\bf c} - $^{232}$Th and {\bf d} -
background (2000 h).}
\end{center}
\end{figure*}

\begin{figure*}[t!]
\begin{center}
\setcaptionmargin{5mm}
\onelinecaptionstrue
\includegraphics[width=10.5cm,angle=0.]{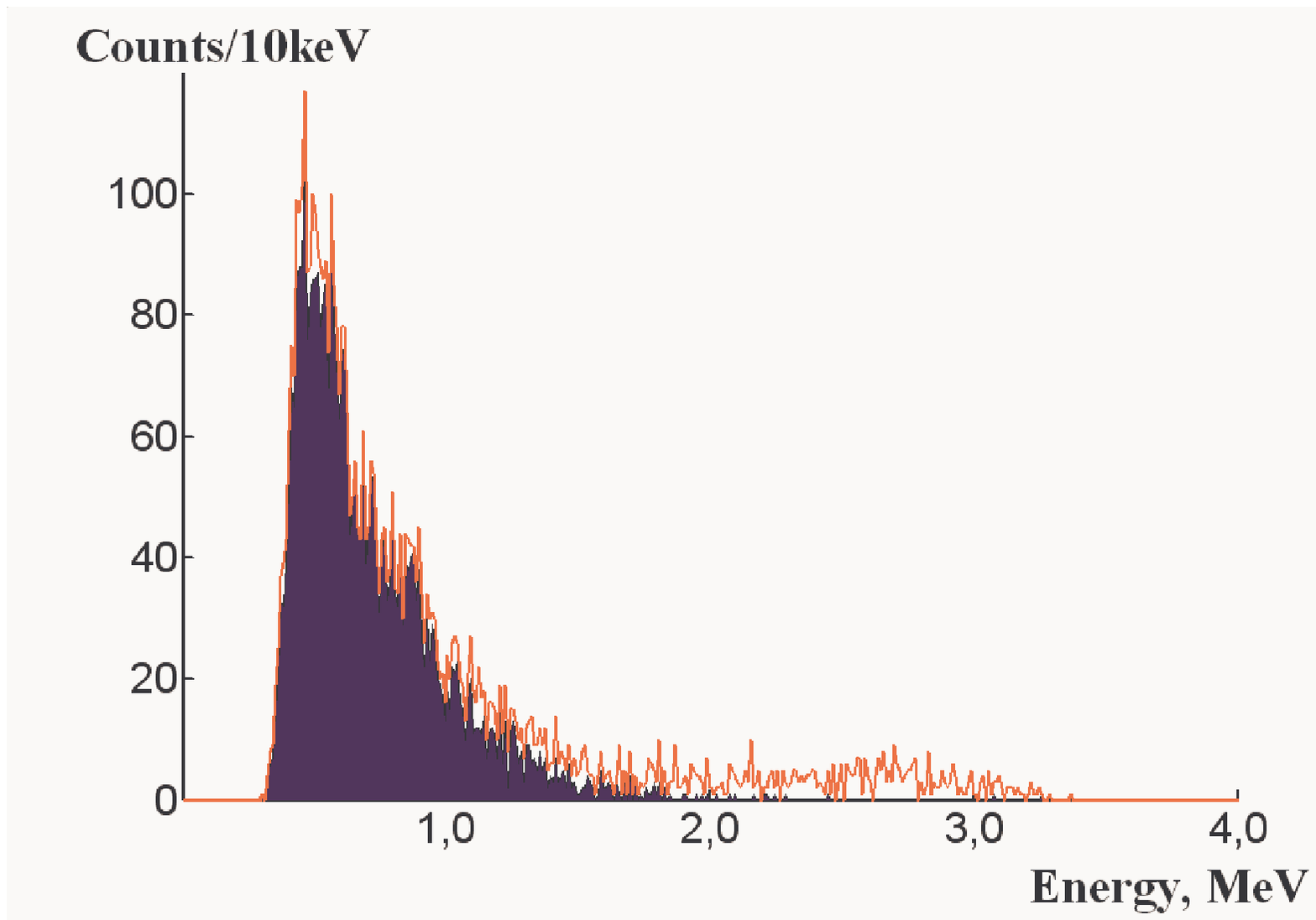}
\captionstyle{normal} \captionstyle{flushleft}
\caption{\label{n_e_sel} The energy spectra measured during 2000 h,
line - before rejection of pulses with $\delta$=100, gray colour
area - after rejection.}
\end{center}
\end{figure*}

\begin{figure*}[t!]
\begin{center}
\setcaptionmargin{5mm}
\onelinecaptionstrue
\includegraphics[width=10.5cm,angle=0.]{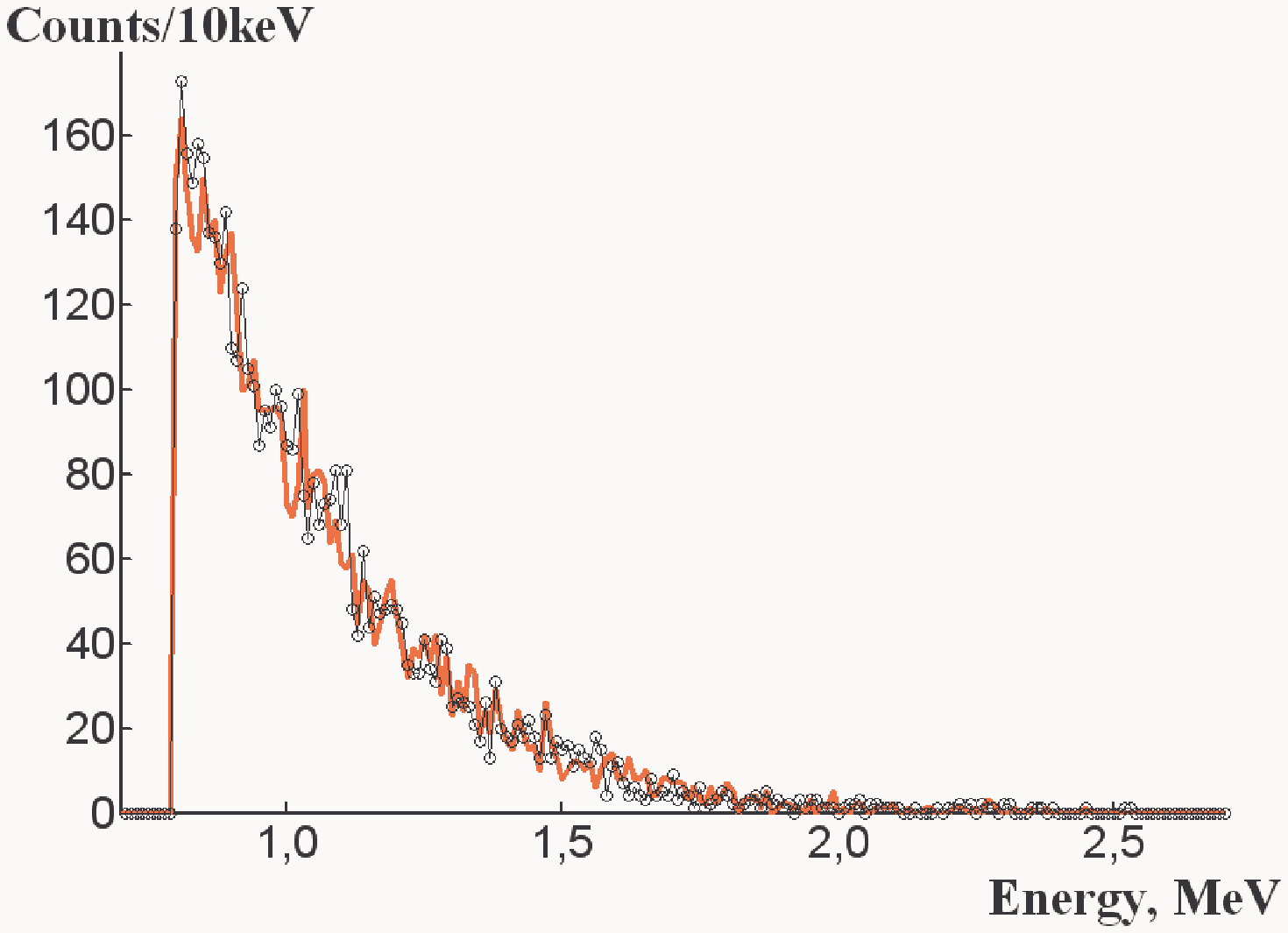}
\captionstyle{normal} \captionstyle{flushleft}
\caption{\label{n_e_tot} The energy spectra measured during 8000 h,
solid line - enriched xenon, open circles - natural xenon.}
\end{center}
\end{figure*}

Distribution of the background (2000 h of measurements), $^{22}$Na,
$^{232}$Th and $^{222}$Rn events versus energy, $\tau$ and $\delta$
are presented in fig.\ref{n_e}, \ref{e_t} and \ref{e_p}. It is seen
that distribution of the $^{232}$Th events differ from $^{222}$Rn
events significantly. This difference is clearly seen in
distribution of events versus energy and $\delta$ for energy greater
then 800keV (see fig.\ref{e_p}.). It was used to separate
$\alpha$-particle events and electron events (pulses from
$\alpha$-particles have $\delta$=100, pulses from electron have
$\delta\leq$100). By rejecting of events with $\delta$=100 we
exclude $\alpha$-particle events (produced in $^{222}$Rn decay chain
in gas volume and coming from internal surface of CPC) from
background spectra. The transformation of the background spectrum
after such rejection is seen in fig.\ref{n_e_sel}. For total
analysis the runs 1,2,4 and 5 were used. The effect value caused by
2$\beta$(2$\nu$)-decay of $^{136}$Xe is determined with comparison
of total spectra measured with enriched and natural xenon (see
fig.\ref{n_e_tot}). Number of events in the energy region
0.8$\div$2.48 MeV registered by CPC $N1$ and CPC $N2$ in each run
during 2000 h of measurements are presented in tab.\ref{tab3}. The
evaluated effect is $-64\pm99(stat)\pm23(syst)$. The total deviation
is $\sigma_{tot}=\sqrt{\sigma^{2}_{stat}+\sigma^{2}_{syst}}=102$. So
effect is $-0.63\sigma_{tot}$. Taking into account efficiencies and
the different percentages of $^{136}$Xe in two gases and the
recommendation given in \cite{r11}, we obtain for the lifetime of
2$\beta(2\nu)$-decay a lower limit:
\begin{eqnarray*}
T_{1/2}(2\beta,2\nu)\geq\frac{\ln(2)\cdot
N_{^{136}Xe}\cdot\epsilon_{1}\cdot\epsilon_{2}}{1.08\cdot\sigma_{tot}}=8.5\cdot10^{21}yr
(90\%\texttt{C.L.})
\end{eqnarray*}
Where: t=8000 h=0.913 $yr$ - measurements time,
$N_{^{\texttt{136}}\texttt{Xe}}=3.21\cdot10^{23}$ - difference in
$^{136}$Xe, $\epsilon_{1}=0.993$ - efficiency after rejection of
events with $\delta=100$ and $\epsilon_{2}=0.467$ - efficiency of
$2\beta(2\nu)$-events registration.

To evaluate the $2\beta(0\nu)$-effect the energy spectra in region
$2317\div2641$ keV were analyzed. This energy region is determined
from recalculated energy resolution for 2479 keV $(R=7.0\%$ or 174
keV) electrons and systematic error in definition of peak position
($\pm$41keV). Number of events in the energy region 2317$\div$2641
keV registered by CPC $N1$ and CPC $N2$ in each run during 2000 h of
measurements are presented in tab.\ref{tab4}. The evaluated effect
is $5-7=-2(\pm4.8)$. Using recommendation given in \cite{r11} and
assuming that mean background is 7 events and measured one is 5
events, we obtain:
\begin{eqnarray*}
T_{1/2}(2\beta,0\nu)\geq\frac{\ln(2)\cdot
N_{^{136}Xe}\cdot\epsilon_{1}\cdot\epsilon_{2}}{3.28}=3.1\cdot10^{23}
yr (90\%\texttt{C.L.})
\end{eqnarray*}
where $\epsilon_{1}=1.0$, $\epsilon_{2}=0.5$

The work was supported by RFBR (grants: 01-02-16069 and
02-02-06136mas).

\newpage

\begin{table}
\setcaptionmargin{0.3mm} \onelinecaptionstrue
\captionstyle{flushleft} \caption{The results of some experiments
for the search of 2$\beta$-decay of $^{136}$Xe} \label{tab1}
\begin{center}
\begin{tabular}{|c|c|c|}
\hline
Experiment & T$_{1/2}$(2$\beta$(2$\nu))$ & T$_{1/2}$(2$\beta$(0$\nu))$\\
\hline Gran Sasso \cite{r1} & $\geq$1.6$\cdot$10$^{20}$yr
(95$\%$C.L.) & $\geq$1.2$\cdot$10$^{22}$yr (95$\%$C.L.)
\\
\hline GOTTHARD \cite{r2} & $\geq$3.6$\cdot$10$^{20}$yr (90$\%$C.L.)
& $\geq$4.4$\cdot$10$^{23}$yr (90$\%$C.L.)
\\
\hline BNO INR RAS \cite{r3} & $\geq$8.1$\cdot$10$^{20}$yr
(90$\%$C.L.) & ...
\\
\hline DAMA/LXe \cite{r4} & $\geq$1.0$\cdot$10$^{22}$yr (90$\%$C.L.)
& $\geq$1.2$\cdot$10$^{24}$yr (90$\%$C.L.)
\\
\hline
\end{tabular}
\end{center}
\end{table}

\begin{table}
\setcaptionmargin{0.3mm} \onelinecaptionstrue
\captionstyle{flushleft} \caption{The theoretical estimations for
2$\beta$(2$\nu$)-decay of $^{136}$Xe}. \label{tab2}
\begin{center}
\begin{tabular}{|c|c|c|} \hline
Authors & T$_{1/2}$(2$\beta$(0$\nu))$\\
\hline E. Caurier et al. \cite{r5} & 2.1$\cdot$10$^{21}$yr
\\
\hline O.A. Rumyantsev, M.G. Urin \cite{r6} & 1.0$\cdot$10$^{21}$yr
\\
\hline A. Staudt et al. \cite{r7} &
1.5$\cdot$10$^{19}$-2.1$\cdot$10$^{22}$yr
\\
\hline P. Vogel and M.R. Zirnbauer \cite{r8} &
1.5$\cdot$10$^{20}$-1.5$\cdot$10$^{21}$yr
\\
\hline
\end{tabular}
\end{center}
\end{table}

\begin{table}
\setcaptionmargin{0.5mm} \onelinecaptionstrue
\captionstyle{flushleft} \caption{The number of events with energy
0.8$\div$2.48MeV, registered in CPC $N1$ and CPC $N2$ during 2000h
of measurements for each run.} \label{tab3} \centering
\begin{center}
\begin{tabular}{|c|c|c|}
\hline
run & CPC $N1$ & CPC $N2$ \\
\hline
1 & 1681 ($^{136}$Xe) & 1108 ($^{nat}$Xe) \\
\hline
2 & 1734 ($^{nat}$Xe) & 1182 ($^{136}$Xe) \\
\hline
4 & 1316 ($^{nat}$Xe) & 783 ($^{136}$Xe) \\
\hline
5 & 1204 ($^{136}$Xe) & 756 ($^{nat}$Xe) \\
\hline
\end{tabular}
\end{center}
\end{table}

\begin{table}
\setcaptionmargin{0.5mm} \onelinecaptionstrue
\captionstyle{flushleft} \caption{The number of events with energy
2317$\div$2641keV, registered in CPC $N1$ and CPC $N2$ during 2000h
of measurements for each run.} \label{tab4} \centering
\begin{center}
\begin{tabular}{|c|c|c|}
\hline
run & CPC $N1$ & CPC $N2$ \\
\hline
1 & 3 ($^{136}$Xe) & 3 ($^{nat}$Xe) \\
\hline
2 & 3 ($^{nat}$Xe) & 1 ($^{136}$Xe) \\
\hline
4 & 1 ($^{nat}$Xe) & 0 ($^{136}$Xe) \\
\hline
5 & 1 ($^{136}$Xe) & 0 ($^{nat}$Xe) \\
\hline
\end{tabular}
\end{center}
\end{table}

\newpage



\begin{thebibliography}{11}

\bibitem{r1} R. Bernabei, P. Belli, F. Cappella et al., Phys. Lett. B{\bf546}, 23 (2002).

\bibitem{r2} R. Luescher, J. Farine, F. Boehm et al. Phys. Lett. B{\bf434}, 407  (1998).

\bibitem{r3} Ju. M. Gavriljuk, V.V. Kuzminov, N. Ya. Osetrova, S.S.
Ratkevich, Phys. Rev. C{\bf61}, 035501 (2000).

\bibitem{r4} R. Bernabei, P. Belli, F. Cappella et al., Phys. Lett. B{\bf546}, 23
(2002).

\bibitem{r5} E. Caurier, F. Nowacki, A. Poves, J. Retamosa, Nucl. Phys. A{\bf654}, 973 (1999).

\bibitem{r6} O.A. Rumyantsev, M.G. Urin, JETP Lett {\bf61}, 361 (1995).

\bibitem{r7} A. Staudt, K. Muto, H. Klapdor-Kleingrothaus, Europhys. Lett. {\bf13}, 31 (1990).

\bibitem{r8} P. Vogel and M.R. Zirnbauer, Phys. Rev. Lett. {\bf57}, 3148 (1986)

\bibitem{r9}  V.N. Gavrin, V.I. Gurentsov,  V.N. Kornoukhov et al, {\it Intensivnost muonov
kosmicheskikh luchei v laboratorii glubokogo zalozheniya GGNT},
Preprint INR RAS, P-698 (1991) Moskva.

\bibitem{r10} Yu. M. Gavriljuk, V.V. Kuzminov, N. Ya. Osetrova, S.S. Ratkevich,
Yad. Phys. {\bf67}, 11, 2039 (2004).

\bibitem{r11} G.J.Feldman, R.D.Causins, Phys.
Rev. D{\bf57}, 3873 (1998).

\end{thebibliography}
\end{document}